\documentclass[10pt, a4paper]{article}
\usepackage{flafter}

\usepackage{amsmath,amssymb}
\usepackage{latexsym}
\usepackage{psfrag}
\usepackage{graphicx,subfigure}

\newtheorem{theorem}{Theorem}[section]

\newtheorem{lemma}[theorem]{Lemma}

\newtheorem{prop}[theorem]{Proposition}


\begin{document}
\date{\today}

\title{Linear sampling method for identifying cavities in a heat conductor}
\author{Horst Heck$^*$, Gen Nakamura$^\dag$ and Haibing Wang$^{\dag,\,\ddag}$
\\${}$\\
$^*$Technische Universit\"at Darmstadt, FB Mathematik\\D-64289 Darmstadt, Germany
\\$^\dag$Department of Mathematics, Hokkaido University\\ Sapporo, 060-0810, Japan
\\$^\ddag$Department of Mathematics, Southeast University\\ Nanjing,
210096, P.R.China\\
{$\begin{array}{l}\mbox{\rm\small E-mail:}\,\,\,\mbox{\rm\small heck@mathematik.tu-darmstadt.de,\, gnaka@math.sci.hokudai.ac.jp,}\\
 \qquad\quad\mbox{\rm\small hbwang@math.sci.hokudai.ac.jp}\end{array}$}}
\maketitle

\begin{abstract}
We consider an inverse problem of identifying the unknown cavities in a heat conductor. Using the Neumann-to-Dirichlet map as an input data, we develop a linear sampling type method for the heat equation. A new feature is that there is a freedom to choose the time variable, which suggests that we have more data than the linear sampling methods for the inverse boundary value problem associated with EIT and inverse scattering problem with near field data.

\end{abstract}
%
%
%
%
%

\section{Introduction}
\setcounter{equation}{0}

The biggest advantage of the linear sampling method for inverse scattering problem is that
it is a very easy numerical scheme to implement. It has attracted and still attracting wide audiences
in inverse scattering problem. Nevertheless, the validation of the scheme
has some problem and it was compensated to some extent by the works of Arens et al. (\cite{Arens1}, \cite{Arens2}).
The linear sampling method
for inverse boundary value problem for EIT (electric impedance tomography) was first given in the unpublished paper by Somersalo (\cite{Somersalo})
to identify unknown inclusions. A similar linear sampling method was considered by Cakoni et al. (\cite{CDS}) for inverse scattering problems using
many near field measurements. In this paper, we will develop a linear sampling type method for the heat equation which could be
of some interest if we not only observe something similar as in the case of inverse boundary value problem and inverse scattering problem, but also
something new. We will also call this method the linear sampling method.

Of course we cannot have scattering problem for the heat equation. A natural inverse problem for
the heat equation would be an inverse boundary value problem which conducts measurements at
the boundary and identify any unknown objects such as inclusions, cavities and cracks. In this paper we will establish a linear sampling method to identify unknown cavities inside a heat conductor by many boundary measurements.

There are many other schemes to identify the unknown objects in heat conductors. We only give some of the references for these schemes. They are the dynamical probe method by Nakamura et al. (\cite{DKN}, \cite{D-Y-L-N2009}, \cite{I-K-N2010}, \cite{KN}, \cite{Y-K-N2010}) and enclosure method by Ikehata et al. (\cite{Ikehata1}, \cite{Ikehata2}, \cite{Ikehata3}, \cite{Ikehata4}).

To begin with let $\Omega\subset{\mathbb R}^n\,\,(n=1,2,3)$ be a heat conductor with
unknown cavities $D\subset\Omega$ which consist of several disjoint domains such that $\overline D\subset\Omega$ and $\Omega\setminus\overline D$ is
connected. Each of these domains can be considered as a cavity. We assume that the boundaries $\partial\Omega$ and $\partial D$ of $\Omega$ and $D$, repectively, are of $C^2$ class.
Let the initial temperature distribution of $\Omega$ with cavities $D$ be zero. By giving a heat flux $f\in L^2((0,\,T);\,H^{-1/2}(\partial\Omega))$
on $\partial\Omega$ over the time interval $(0,\,T)$, we measure the corresponding temperature distribution on $\partial\Omega$ over $(0,\,T)$ by
an infrared light camera which is the case for thermography.
If we can repeat this measurement so many times so that we can almost have the Neumann-to-Dirichlet map
\begin{equation}\label{eq:ND0}
\Lambda_D:\,L^2((0,\,T);\, H^{-1/2}(\partial\Omega))\rightarrow L^2((0,\,T);\,H^{1/2}(\partial\Omega))
\end{equation}
which maps the heat flux $f$ on $\partial\Omega$ over the time interval $(0,\,T)$ to the corresponding temperature
distribution of $\partial\Omega$ over $(0,\,T)$.
This can be really the case
if the given heat fluxes are confined at a part of $\partial\Omega$ and we have the cooling boundary condition at
the other part of the boundary, because in this case the temperature generated by the heat flux goes to $0$ exponentially fast in time
after we stop giving the heat flux. To make our argument simple, we will remain to the case there is no
cooling boundary condition on the part of $\partial\Omega$.
Then, our inverse boundary value problem is
as follows.

\bigskip
We define more precisely the Neumann-to-Dirichlet map $\Lambda_D$ by formulating a forward problem whose solution describes
how the temperature distribution $u$ in $\Omega\setminus\overline D$ generated by the heat flux $f$ diffuses. To begin with let
{\small
$$W((\Omega\setminus\overline D)_T):=\left\{  u\in L^2((0,\,T);\, H^1(\Omega\setminus\overline D)),\,\partial_t u\in L^2( (0,\,T);\,(H^1(\Omega\setminus\overline D))^\ast=\dot{H}^{-1}( \overline{\Omega\setminus\overline D} ) ) \right\}.$$}
Then, the \textit{forward problem} is to look for a function $u=u^f=u(f)\in W((\Omega\setminus\overline D)_T)$
which satisfies
\begin{equation}\label{eq:mp}
\left\{\begin{array}{l}
(\partial_t - \Delta)u=0\textrm{ in }(\Omega\setminus\overline D)\times (0,\,T)=:(\Omega\setminus\overline D)_T, \\
\partial_\nu u=f\in L^2( (0,\,T); H^{-1/2}(\partial\Omega) )\textrm{ on }\partial \Omega\times (0,\,T)=: (\partial \Omega)_T,\\
\partial_\nu u=0\textrm{ on }\partial D\times (0,\,T)=: (\partial D)_T, \\
u=0\textrm{ at } t=0.
\end{array}\right.
\end{equation}
Here $\nu$ is the unit normal vector to $\partial \Omega\,(\textrm{or }\partial D)$ directed outside $\Omega \, (\textrm{or }D)$
and for any set $E\subset{\mathbb R}^n$, $E_T$ denotes the cylindrical set $E\times (0,\,T)$.

This is a well-posed problem, by which we can define the Neumann-to-Dirichlet map $\Lambda_D$:
\begin{equation}\label{eq:ND}
\Lambda_D:\, L^2((0,\,T);H^{-1/2}(\partial\Omega))\to L^2((0,\,T); H^{1/2}(\partial\Omega)),\, f\mapsto u^f|_{(\partial\Omega)_T}.
\end{equation}
If there is no any cavity, i.e. $D=\emptyset$, we denote the Neumann-to-Dirichlet map by $\Lambda_\emptyset$.

When we take the Neumann-to-Dirichlet map $\Lambda_D$ as the measured data, our inverse boundary value problem is formulated as follows:

\medskip
{\bf Inverse Problem:} Reconstruct $D$ from $\Lambda_D$.

\medskip
The linear sampling method considers some integral equation $Fg=\Gamma_{(y,s)}^0$ of the first kind for $F=:\Lambda_D-\Lambda_\emptyset$, where $\Gamma_{(y,s)}^0$ is the Green function of the heat operator $\partial_t-\Delta$ in $\Omega\times (0,T)$ with the Neumann boundary condition on $\partial\Omega\times(0,T)$ which has a singularity at $(y,s)$. Then, the linear sampling method
examines the behavior of an approximation solution to this integral equation for the case $y\in D$ and $y\not\in D$ when we fix $s$. A new feature which we can observe here is that there is a freedom to choose $s$, which suggests that we have more data than the linear sampling methods for the inverse boundary value problem for EIT and inverse scattering problem with near field data.

The rest of this paper is organized as follows. In the next section, we will give the linear sampling method for the case of $y\in D$ and $y\not\in D$, respectively. Some properties of the heat potential used in Section 2 will be proved in Appendix.

\section{Linear sampling method for identifying the cavity}
\setcounter{equation}{0}

Let
$$\Gamma_{(y,\,s)}(x,\,t):=\Gamma(x,\, t;\, y,\, s)=
\left\{ \begin{array}{ll}\frac{1}{(4\pi (t-s))^{n/2}} \exp\left( -\frac{| x-y |^2}{4(t-s)} \right),& t>s,\\
0,& t\leq s
\end{array}\right.
$$
be the fundamental solution to the heat operator $\partial_t -\Delta$, and $\Gamma^0_{(y,\,s)}(x,\,t)$ be the Green function of the heat operator with Neumann boundary condition on $(\partial\Omega)_T$. We define the operators $S,\,H,\,A,\,F$ as follows:
\begin{itemize}
\item $Sf:=u^f,\,Hf:=\partial_\nu Sf|_{(\partial D)_T}$ with the solution $u^f$ to
\begin{equation*}
\left\{ \begin{array}{l}
(\partial_t - \Delta)u^f=0\textrm{ in }\Omega_T,\\
\partial_\nu u^f =f \textrm{ on }(\partial\Omega)_T,\\
u^f=0 \textrm{ at } t=0;
\end{array}\right.
\end{equation*}

\item $Ag:=z^g|_{(\partial\Omega)_T}$ with the solution $z^g$ to
\begin{equation*}
\left\{ \begin{array}{l}
(\partial_t - \Delta)z^g=0\textrm{ in }(\Omega\setminus\overline D)_T,\\
\partial_\nu z^g|_{(\partial D)_T}=g,\,\partial_\nu z^g|_{(\partial \Omega)_T}=0,\\
z^g|_{t=0}=0;
\end{array}\right.
\end{equation*}

\item $F:=\Lambda_D - \Lambda_{\emptyset}$.
\end{itemize}
Then, we have the following lemmas:

\begin{lemma}\label{lem:1}
$F=-AH$.
\end{lemma}
{\bf Proof.} Let $v=Sf$ with $f\in L^2((0,\,T); H^{-1/2}(\partial\Omega))$ and consider the solutions $w$ and $u$ to
\begin{equation*}
\left\{ \begin{array}{l}
(\partial_t - \Delta)w=0\textrm{ in }(\Omega\setminus\overline D)_T,\\
\partial_\nu w|_{(\partial D)_T}=-\partial_\nu v|_{(\partial D)_T},\, \partial_\nu w|_{(\partial \Omega)_T} =0,\\
w|_{t=0}=0
\end{array}\right.
\end{equation*}
and
\begin{equation*}
\left\{ \begin{array}{l}
(\partial_t - \Delta)u=0\textrm{ in }(\Omega\setminus\overline D)_T,\\
\partial_\nu u|_{(\partial D)_T}=0,\, \partial_\nu u|_{(\partial \Omega)_T} = \partial_\nu v|_{(\partial \Omega)_T},\\
u|_{t=0}=0,
\end{array}\right.
\end{equation*}
respectively.
Since $(\partial_t - \Delta) v=0$ in $(\Omega\setminus\overline D)_T$, we have
\begin{equation*}
\left\{ \begin{array}{l}
(\partial_t - \Delta)(u-v)=0\textrm{ in }(\Omega\setminus\overline D)_T,\\
\partial_\nu (u-v)|_{(\partial D)_T}=\partial_\nu w|_{(\partial D)_T},\, \partial_\nu (u-v)|_{(\partial \Omega)_T} =0,\\
(u-v)|_{t=0}=0,
\end{array}\right.
\end{equation*}
which implies $w=u-v$ by the uniqueness. Hence, by $\partial_\nu v|_{(\partial D)_T}=Hf$, we have
\begin{eqnarray*}
A(-Hf)&=&w|_{(\partial\Omega)_T}=(u-v)|_{(\partial \Omega)}\\
&=&(\Lambda_D - \Lambda_\emptyset)(\partial_\nu v|_{(\partial\Omega)_T})=(\Lambda_D - \Lambda_\emptyset)f=Ff.
\end{eqnarray*}
Therefore, $F=-AH$. \hfill $\Box$

\begin{lemma}\label{lem:2}
The operator $$H:\, L^2( (0,\,T);\, H^{-1/2}(\partial\Omega) )\to L^2( (0,\,T);\, H^{-1/2}(\partial D) )$$ is continuous and has dense range.
\end{lemma}
{\bf Proof.} We prove the result by using layer potential argument for the heat equation. To this end, we introduce the function space $H^{r,s}$ defined by
$$H^{r,s}(\Gamma_T):=L^2((0,\,T);\,H^r(\Gamma))\cap H^s((0,\,T);\,L^2(\Gamma)),\quad \Gamma=\partial D \textrm{ or } \partial \Omega.$$
Recall that $u:=S f$ satisfies
\begin{equation}\label{eq:den1}
\left\{ \begin{array}{l}
(\partial_t - \Delta)u=0\textrm{ in }\Omega_T,\\
\partial_\nu u =f \textrm{ on }(\partial\Omega)_T,\\
u=0 \textrm{ at } t=0.
\end{array}\right.
\end{equation}
We express the solution $u$ to (\ref{eq:den1}) as a single-layer heat potential:
\begin{equation}\label{eq:den2}
u(x,\,t)=K_0\phi=:\int_0^t \int_{\partial \Omega} \Gamma(x,\,t;\,y,\,s) \phi(y,\,s)ds(y)ds,\quad (x,\,t)\in \Omega_T
\end{equation}
with an unknown density $\phi\in H^{\frac{1}{2},\frac{1}{4}}((\partial\Omega)_T)$. Define the operator $N$ by
$$ N\phi := \frac{1}{2} \left[ \gamma_1 (K_0 \phi)|_{\Omega} +  \gamma_1 (K_0 \phi)|_{\Omega^c}\right],  $$
where $$\gamma_1:\,H^{2,1}(\Omega_T)\to H^{\frac{1}{2},\frac{1}{4}}((\partial\Omega)_T)$$ is a continuous linear map \cite{Costabel1990}.
Then, using the jump relations of layer potentials in Theorem 3.7 of \cite{Costabel1990}, we derive that
\begin{equation}\label{eq:den3}
\frac{1}{2}\phi+N\phi = f.
\end{equation}
According to Proposition 4.3 in \cite{Costabel1990}, we know that
\begin{equation}\label{eq:den4}
\frac{1}{2}I+N:\,H^{\frac{1}{2},\frac{1}{4}}((\partial\Omega)_T) \to H^{\frac{1}{2},\frac{1}{4}}((\partial\Omega)_T)
\end{equation}
is an isomorphism. By defining $$K\psi:=\partial_\nu (K_0\psi)|_{(\partial D)_T}$$ for $\psi\in L^2((0,\,T);\,H^{-\frac{1}{2}}(\partial\Omega))$, we have $$H=K(\frac{1}{2}+N)^{-1}.$$ In addition, the well-posedness of the forward problem says that the operator
\begin{equation}\label{eq:den5}
H:\, L^2( (0,\,T);\, H^{-1/2}(\partial\Omega) )\to L^2( (0,\,T); H^{-1/2}(\partial D) )
\end{equation}
is continuous. Note that $H^{\frac{1}{2},\frac{1}{4}}((\partial \Omega)_T)$ is dense in $L^2((0,\,T);\,H^{-\frac{1}{2}}(\partial \Omega))$. It is enough to show that the operator
$$K:\,L^2((0,\,T);\,H^{-\frac{1}{2}}(\partial\Omega))\to L^2((0,\,T);\,H^{-\frac{1}{2}}(\partial D))$$
has dense range. We give the proof as follows.

By direct calculations, we have that
\begin{equation}\label{eq:den6}
K\psi(x,\,t)=\int_0^t \int_{\partial\Omega}M(x,\,t;\,y,\,s)\psi(y,\,s)ds(y)ds,\quad (x,\,t)\in (\partial D)_T
\end{equation}
with
$$ M(x,\,t;\,y,\,s):=\partial_{\nu(x)} \Gamma(x,\,t;\,y,\,s).$$
Let $$K^*:\,L^2((0,\,T);\,H^{\frac{1}{2}}(\partial D))\to L^2((0,\,T);\,H^{\frac{1}{2}}(\partial \Omega))$$ be the adjoint of $K$. It follows that
\begin{equation}\label{eq:den7}
K^*\eta(y,\,s)=\int_s^T \int_{\partial D}M(x,\,t;\,y,\,s)\eta(x,\,t)ds(x)dt,\quad (y,\,s)\in (\partial \Omega)_T.
\end{equation}
For proving the denseness of $K$, it suffices to show that $\eta=0$ if $$ (K\psi,\,\eta)=(\psi,\,K^*\eta)=0 $$ for all $\psi\in L^2((0,\,T);\,H^{-\frac{1}{2}}(\partial \Omega))$.

Define
\begin{equation}\label{eq:den8}
w(y,\,s)=\int_s^T \int_{\partial D}M(x,\,t;\,y,\,s)\eta(x,\,t)ds(x)dt,\quad (y,\,s)\in (\mathbb{R}^m\setminus\overline D)_T.
\end{equation}
According to Appendix \ref{Appendix:uniqueness}, we have
$$w=0 \textrm{ in } (\mathbb{R}^n\setminus\overline{D})_T,$$
whence $$w=0 \textrm{ on } (\partial D)_T.$$ By the uniqueness result for the heat equation in $D_T$, we also have
$$w=0 \textrm{ in } D_T.$$
Consequently, we obtain that $\eta=0$ from
the jump formula of the operator $K^*$ as $y$ tends to $\partial D$, which is shown in Appendix \ref{Appendix:jump}. The proof is complete. \hfill  $\Box$

\bigskip
We are now in a position to state the main result for the linear sampling method:
\begin{prop}\label{prop:3}
Fix $s\in(0,\,T)$ and let $y\in D$. Then, there exists a function $g^y\in L^2( (0,\,T); H^{-1/2}(\partial\Omega))$ satisfying the inequality
\begin{equation}\label{eq:data1}
\| Fg - \Gamma^0_{(y,\,s)} \|_{L^2( (0,\,T);\, H^{1/2}(\partial\Omega))}< \varepsilon
\end{equation}
such that
\begin{equation}\label{eq:density11}
\lim_{y\to\partial D} \| g^y \|_{L^2( (0,\,T);\, H^{-1/2}(\partial\Omega))}=\infty
\end{equation}
and
\begin{equation}\label{eq:density12}
\lim_{y\to\partial D} \| Hg^y \|_{L^2( (0,\,T);\, H^{-1/2}(\partial D))}=\infty.
\end{equation}
\end{prop}
{\bf Proof.} We will adopt the proof given in \cite{C-C2006} and \cite{Pot2001}, which is for the scattering case. Let $\| A \|$ be the norm of the operator $$A:\,L^2( (0,\,T);\, H^{-1/2}(\partial D)) \to L^2( (0,\,T);\, H^{1/2}(\partial\Omega)).$$ For given $\varepsilon > 0$, by Lemma \ref{lem:2}, there exists a function $g^y\in L^2( (0,\,T);\, H^{-1/2}(\partial\Omega))$ such that
\begin{equation*}
\| Hg^y - (-\partial_\nu \Gamma^0_{(y,\,s)}) \|_{L^2( (0,\,T);\, H^{-1/2}(\partial D))}< \frac{\varepsilon}{\| A \|}.
\end{equation*}
Since $\partial_\nu \Gamma^0_{(y,\,s)}|_{(\partial\Omega)_T}=0$, we have
$$
A( \partial_\nu \Gamma^0_{(y,\,s)}|_{(\partial D)_T} )=\Gamma^0_{(y,\,s)}|_{(\partial\Omega)_T}.
$$
Hence, by Lemma \ref{lem:1}, we have
\begin{eqnarray*}
&&\| F g^y - \Gamma^0_{(y,\,s)} \| _{L^2( (0,\,T);\, H^{1/2}(\partial\Omega))}\\
&=& \| -AHg^y - A (\partial_\nu \Gamma^0_{(y,\,s)}|_{(\partial D)_T}) \|_{L^2( (0,\,T);\, H^{1/2}(\partial\Omega))}\\
&\leq & \| A \| \, \| H g^y -(-\partial_\nu \Gamma^0_{(y,\,s)}|_{(\partial D)_T}) \|_{L^2( (0,\,T);\, H^{-1/2}(\partial D))} < \varepsilon.
\end{eqnarray*}
On the other hand, due to the boundedness of $H$, we have
\begin{eqnarray*}
&&\| g^y \|_{L^2( (0,\,T);\, H^{-1/2}(\partial\Omega))}\\
& \geq & c \| H g^y \|_{L^2( (0,\,T);\,H^{-1/2}(\partial D))}\\
&\geq & c \left( \| \partial_\nu \Gamma^0_{(y,\,s)}\|_{L^2( (0,\,T);\, H^{-1/2}(\partial D))}- \| H g^y- (-\partial_\nu \Gamma^0_{(y,\,s)})\|_{L^2( (0,\,T);\, H^{-1/2}(\partial D))} \right)\\
&\geq & c \| \partial_\nu \Gamma^0_{(y,\,s)}\|_{L^2( (0,\,T);\, H^{-1/2}(\partial D))} - \frac{c\varepsilon}{\| A \|}
\end{eqnarray*}
with some constant $c>0$.

\medskip
We next prove $$\| \partial_\nu \Gamma^0_{(y,\,s)} \|_{L^2( (0,\,T);\, H^{-1/2}(\partial D))}\to \infty\quad (y\to \partial D).$$

We only consider the case of $n=3$ (for $n=1,\,2$, the proof is much easier). By the assumption on $\partial D\in C^2$, for any point $x_0\in\partial D$ there exists a $C^2$-function $f$ such that $$D\cap B(x_0,\, r)=\left\{ x\in B(x_0,\, r):\,x_3> f(x_1,\, x_2) \right\},$$
where $B(x_0,\, r)$ is the ball with radius $r>0$ and center at $x_0$. We choose a new orthonormal basis $\{e_j\},\,j=1,\,\cdots,\,3,$ centered at $x_0$ with $e_3=-\nu$, where $\nu$ is the normal unit vector to the boundary at $x_0$. The vectors $e_1,\,e_2$ lie in the tangent plane to $\partial D$ at $x_0$. Let $\eta$ be the local coordinates defined by the basis $\{e_j\}$. Define the local transformation of coordinates $\eta=F(x)$ as follows:
\begin{equation*}
\eta^\prime=x^\prime,\,\eta_3=x_3-f(x^\prime) \textrm{ with } x^\prime=(x_1,\,x_2),\,\eta^\prime=(\eta_1,\,\eta_2).
\end{equation*}
Thus, in the new coordinates system, we can let $\xi=(0,\,0,\,\xi_3),\,\xi_3>0$ and $\eta^\prime\in [-l,\, l]\times [-l,\, l]:=D_2$ with a sufficiently small constant $l$. Without loss of generality, we set $s=0$. Then, we have
\begin{eqnarray*}
& &\| \partial_\nu \Gamma^0_{(y,\,s)}\|_{L^2( (0,\,T);\, H^{-1/2}(\partial D))} = \int^T_0 \| \partial_\nu \Gamma^0_{(y,\,s)} \|^2_{H^{-1/2}(\partial D)}dt \\
&\geq & c_1\int^T_0 \| \partial_\nu \Gamma^0_{(\xi^\prime,\,s)} \|^2_{H^{-1/2}(D_2)}dt \\
&=&c_1 \int^T_0 \left\{ \sup_{\|\varphi\|_{H^{1/2}(D_2)}\leq 1} \left|\int_{D_2}  \partial_\nu \Gamma^0_{(\xi^\prime,\,s)}(\eta^\prime,\,t)\varphi(\eta^\prime)d\eta^\prime \right|  \right\}^2 dt\\
&\geq &c_1 \int^T_0 \left\{ \sup_{\|\varphi\|_{H^1(D_2)}\leq 1} \left| \int_{D_2} \partial_\nu \Gamma^0_{(\xi^\prime,\,s)}(\eta^\prime,\,t)\varphi(\eta^\prime)d\eta^\prime \right|  \right\}^2 dt,
\end{eqnarray*}
where $c_1$ is a constant.

We now estimate the last term in above inequality. To this end, we introduce an auxiliary function $\varphi(\eta^\prime)=ce^{-\frac{| \eta^\prime |^2}{4t}}$. By direct calculations, we obtain that
\begin{equation*}
\int_{D_2} \varphi^2(\eta^\prime) d\eta^\prime \leq c^2 \int_{\mathbb R^2} e ^{-\frac{| \eta^\prime |^2}{2t}} d \eta^\prime = 2c^2 t\int_{\mathbb R^2} e^{-| \xi^\prime |^2} d\xi^\prime = 2c^2 \sqrt\pi t,
\end{equation*}
\begin{equation*}
\int_{D_2} (\partial_{\eta_j} \varphi (\eta^\prime) )^2 d\eta^\prime = c^2 \frac{1}{4t^2}\int_{D_2} \eta^2_j e^{-\frac{| \eta^\prime |^2}{2t}}d\eta^\prime.
\end{equation*}
Hence, we deduce that
\begin{eqnarray*}
&&\int_{D_2} | \nabla_{\eta^\prime} \varphi(\eta^\prime) |^2 d\eta^\prime \leq c^2 \frac{1}{4t^2}\int_{\mathbb R^2} | \eta^\prime |^2 e^{-\frac{| \eta^\prime |^2}{2t}}d\eta^\prime = c^2 \int_{\mathbb R^2} | \gamma^\prime |^2 e^{-| \gamma^\prime |^2} d\gamma^\prime\\
&=& c^2 \int_{\mathbb{R}^2} | \gamma^\prime |^2 e^{-| \gamma^\prime |^2}d\gamma^\prime = 2\pi c^2 \int^\infty_0 r^3 e^{-r^2} dr =2\pi c^2 \int^\infty_0 r e^{-r^2} dr\\
&=& 2\pi c^2 \left[ -\frac{1}{2} e^{- r^2} \right]^\infty_0 = \pi c^2,
\end{eqnarray*}
and therefore
\begin{equation*}
\| \varphi \|^2_{H^1(D_2)} \leq c^2 \left( 2\sqrt\pi t + \pi \right) \leq c^2  \left( 2\sqrt\pi T + \pi \right).
\end{equation*}
In the following, we take $c$ to satisfy $$c^2 \left( 2\sqrt\pi T + \pi \right)\leq 1.$$
Note that
$$ \nu(\eta)\cdot (\eta-\xi) = (0,\,0,\,-1)\cdot(\eta^\prime,\, -\xi_3) = \xi_3,\quad | \eta-\xi |^2 = | \eta^\prime |^2 + \xi^2_3.$$
We have
\begin{eqnarray*}
\partial_\nu \Gamma^0_{(\xi,\,s)}(\eta,\,t)&=&\frac{1}{(\sqrt{4\pi t})^3} \frac{-\nu(\eta)\cdot (\eta-\xi)}{2t} \exp\left(-\frac{| \eta-\xi |^2}{4t}\right)\\
&=& -\frac{1}{16\pi^{3/2}}t^{-5/2}\xi_3 \exp\left( -\frac{| \eta^\prime |^2 + \xi_3^2}{4t} \right).
\end{eqnarray*}
Taking $\varphi=ce^{-\frac{| \eta^\prime |^2}{4t}}$ yields
\begin{eqnarray*}
 &&\sup_{|\varphi|_{H^1(D_2)}\leq 1} \left| \int_{D_2} \partial_\nu \Gamma^0_{(\xi^\prime,\,s)}(\eta^\prime,\,t)\varphi(\eta^\prime)d\eta^\prime \right| dt\\
 &\geq& \frac{c}{16\pi^{3/2}} t^{-5/2} \xi_3 e^{-\frac{\xi^2_3}{4t}} \int_{D_2} e^{-\frac{| \eta^\prime |^2}{2t}} d\eta^\prime \\
 &=& \frac{c}{8\pi^{3/2}} t^{-3/2} \xi_3 e^{-\frac{\xi^2_3}{4t}} \int_{D_t} e^{-| \gamma^\prime |^2} d\gamma^\prime
 \end{eqnarray*}
with $$D_t=[-\frac{l}{\sqrt{2t}},\,\frac{l}{\sqrt{2t}}]\times [-\frac{l}{\sqrt{2t}},\,\frac{l}{\sqrt{2t}}].$$ If we note that $$I(t):=\int_{D_t} e^{-| \gamma^\prime |^2} d\gamma^\prime$$ is monotonically decreasing in $[0,\, T]$, it can be concluded that
\begin{eqnarray*}
&&\| \partial_\nu \Gamma^0_{(y,\,s)}\|^2_{L^2( (0,\,T); H^{1/2}(\partial D))} \\
&\geq& \frac{c^2c_1^2I^2(T)}{64\pi^3} \xi^2_3 \int^T_0 t^{-3} e^{-\frac{\xi^2_3}{2t}}dt\\
&=& \frac{c^2c_1^2I^2(T)}{64\pi^3} \xi^{-2}_3 \int^\infty_{\frac{\xi_3^2}{T}} s e^{-\frac{s}{2}}ds \to \infty\quad (\xi_3\to 0 \textrm{ as } y\to\partial D).
\end{eqnarray*}
The proof is complete. \hfill $\Box$

\bigskip
To further investigate the behavior of the density $g^y$ as $y$ approaches to $\partial D$ from the outside of $D$, we need the following two lemmas.
\begin{lemma}\label{lem:4}
The operator
$$A:\,L^2( (0,\,T);\, H^{-1/2}(\partial D) ) \to L^2( (0,\,T);\, H^{1/2}(\partial\Omega) )$$
is injective and has dense range.
\end{lemma}
{\bf Proof.} The injectivity immediately follows from the unique continuation property for $\partial_t - \Delta$. For the denseness, we will adopt the proof of Lemma 4.1 given in \cite{C-L-N2003}, which is for the scattering case. Let $g_j\in L^2( (0,\,T);\, H^{-1/2}(\partial D) )\,(j\in\mathbb{N})$ be such that the linear hull $[\{g_j\}]$ of $\{g_j\}$ is dense in $L^2( (0,\,T);\, H^{-1/2}(\partial D) )$. By Hahn-Banach's theorem, it is enough to prove the following:
$$
f\in L^2( (0,\,T); H^{-1/2}(\partial \Omega) ):\, \int_{(\partial \Omega)_T}\varphi_j f dsdt =0 \textrm{ for all }j\in \mathbb{N}\textrm{ with }\varphi_j:=z^{g_j}|_{(\partial \Omega)_T}
$$
implies $f=0$.

Consider the solution $v\in W((\Omega\setminus\overline D)_T)$ to
\begin{equation*}
\left\{ \begin{array}{l}
(\partial_t + \Delta)v=0\textrm{ in }(\Omega\setminus\overline D)_T,\\
\partial_\nu v|_{(\partial D)_T}=0,\, \partial_\nu v|_{(\partial \Omega)_T} = f,\\
v|_{t=T}=0,
\end{array}\right.
\end{equation*}
and set $z_j=z^{g_j}$. Then, we have
\begin{eqnarray*}
0&=&\int_{(\Omega\setminus\overline D)_T} ( v \Delta z_j - z_j \Delta v )dxdt \\
&=& \int_{(\partial\Omega)_T} ( \partial_\nu z_j v - \partial_\nu v z_j )dsdt - \int_{(\partial D)_T} ( \partial_\nu z_j v - \partial_\nu v z_j )dsdt \\
&=& -\int_{(\partial D)_T} g_j v dsdt.
\end{eqnarray*}
Hence $v=0$ on $(\partial D)_T$. Using this fact together with $\partial_\nu v|_{(\partial D)_T}=0$, we have $v=0$ in $(\overline\Omega\setminus\overline D)_T$. Therefore $f=\partial_\nu v|_{(\partial \Omega)_T}=0$. The proof is complete.
\hfill $\Box$

\begin{lemma}\label{lem:5}
For any fixed $s\in(0,\, T)$, $\Gamma^0_{(y,\, s)}\not\in R(A):=\textrm{ range of } A$ if $y\in \Omega\setminus D$.
\end{lemma}
{\bf Proof.} Suppose $\Gamma^0_{(y,\, s)}\in R(A)$. Then there exist a function $f\in L^2( (0,\,T);\, H^{-1/2}(\partial D) )$ and the solution $w^f\in W((\Omega\setminus\overline D)_T)$ to
\begin{equation*}
\left\{ \begin{array}{l}
(\partial_t - \Delta) w^f=0\textrm{ in }(\Omega\setminus\overline D)_T,\\
\partial_\nu w^f |_{(\partial D)_T}=f,\,\partial_\nu w^f |_{(\partial\Omega)_T}=0,\\
w^f|_{t=0}=0
\end{array}\right.
\end{equation*}
such that
$$ w^f|_{(\partial\Omega)_T}=\Gamma^0_{(y,\,s)}. $$
Since $\partial_\nu w^f|_{(\partial\Omega)_T}=\partial_\nu \Gamma^0_{(y,\,s)}|_{(\partial\Omega)_T}=0$, we have by the unique continuation principle
\begin{equation*}
w^f=\Gamma^0_{(y,\,s)} \textrm{ in }\left(\overline\Omega\setminus(\overline D \cup \{y\})\right)_T
\end{equation*}
and hence
\begin{equation*}
\| \Gamma^0_{(y,\,s)} \|_{ L^2( (0,\,T);\, H^1(\Omega\setminus\overline D) ) } =\| w^f \|_{L^2( (0,\,T);\, H^1(\Omega\setminus\overline D) )}<\infty.
\end{equation*}
Here $y\in \Omega\setminus D$ can be either $y\in \partial D$ or $y\in\Omega\setminus\overline D$, but which ever the case it holds
\begin{equation}\label{eq:Gamma}
\| \Gamma^0_{(y,\, s)} \|_{ L^2( (0, \,T);\, H^1(\Omega\setminus\overline D) ) } =\infty.
\end{equation}

\medskip
In the following, we give a proof for (\ref{eq:Gamma}).

Let $n=3,\,s=0$ and $y\in \partial D$ (The other case can be shown more easily). Then, for some $\delta\,(0<\delta<\pi/2),\,\gamma\, (| \gamma | <\pi),\,R,\,\tau\,(0<R,\,\tau\ll 1)$ and any $\varepsilon\, (0<\epsilon\ll 1)$, we have
\begin{eqnarray*}
&&\left(\frac{1}{8\pi^{3/2}} \right)^2 \int^T_0 \int_{\Omega\setminus\overline D} t^{-3} \exp\left( -\frac{| x-y |^2}{2t} \right)dxdt\\
&\geq & \left(\frac{1}{8\pi^{3/2}} \right)^2 \int^\tau_\varepsilon dt \int^{\pi-\delta}_\delta d\varphi \int^{\gamma}_{-\gamma} d\theta \int^R_0 t^{-3}\exp\left( -\frac{r^2}{2t} \right) r^2 \sin\varphi dr\\
&=& \frac{1}{64\pi^3} [-\cos\varphi]^{\pi-\delta}_{\delta} 2\gamma \int^\tau_\epsilon t^{-3} dt \int^R_0 r^2 \exp\left( -\frac{r^2}{2t} \right) dr\\
&\geq & \frac{1}{16\pi^3} \gamma \cos\delta \int^\tau_\varepsilon t^{-1} dt \to \infty \quad (\varepsilon \to 0).
\end{eqnarray*}
Therefore
\begin{equation*}
\| \Gamma^0_{(y,\, s)} \|_{ L^2( (0, \,T); L^2(\Omega\setminus\overline D) ) } =\infty
\end{equation*}
and hence
\begin{equation*}
\| \Gamma^0_{(y,\,s)} \|_{ L^2( (0, \,T); H^1(\Omega\setminus\overline D) ) } =\infty.
\end{equation*}
This is a contradiction.  \hfill $\Box$

\bigskip
In contrast to Proposition \ref{prop:3} for $y\in D$, we now establish the following blowup property of the density $g^y$ for $y\not\in D$ as follows:
\begin{prop}\label{prop:6}
Fix $s\in (0,\, T)$  and let $y\in \Omega\setminus D$. Then, for every $\varepsilon > 0$ and $\delta>0$, there exists $g^y_{\varepsilon,\,\delta}=g^y\in L^2( (0,\,T);\, H^{-1/2}(\partial\Omega))$ satisfying the inequality
\begin{equation}\label{eq:data2}
\| Fg^y - \Gamma^0_{(y,\,s)} \|_{L^2( (0,\, T);\, H^{1/2}(\partial\Omega))}< \varepsilon+\delta
\end{equation}
such that
\begin{equation}\label{eq:blowup1}
\lim_{\delta\to 0} \| g^y \|_{L^2( (0,\,T);\, H^{-1/2}(\partial\Omega)) }=\infty
\end{equation}
and
\begin{equation}\label{eq:blowup2}
\lim_{\delta\to 0} \| Hg^y \|_{L^2( (0,\, T);\, H^{-1/2}(\partial D)) }=\infty.
\end{equation}
\end{prop}
{\bf Proof.} We will adopt the argument given in \cite{C-C2006} and \cite{Pot2001}, which is used for the scattering case.
Since $A$ is injective and has a dense range, for arbitrary $\delta>0$ there exists a function
$$f^z_\alpha = \sum \frac{\mu_n}{\alpha+\mu_n^2}(\Gamma^0_{(y, s)},\, g_n) \varphi_n$$
with the singular system $(\mu_n,\, \varphi_n,\, g_n)$ of $A$
such that
\begin{equation*}
\| A f^z_\alpha - \Gamma^0_{(y,\, s)} \|_{ L^2( (0,\, T);\, H^{1/2}(\partial\Omega)) }< \delta.
\end{equation*}
By Lemma \ref{lem:5}, we have $\Gamma^0_{(y,\, s)}\not\in R(A)$. Hence, by Picard's theorem, we have
 \begin{equation*}
\| f^z_\alpha \|_{ L^2( (0,\, T);\, H^{-1/2}(\partial D)) }\to\infty \quad (\alpha\to 0).
\end{equation*}
By the denseness of $R(H)$ in $ L^2( (0,\, T);\, H^{-1/2}(\partial D))$, there exists $$g^z_\alpha\in  L^2( (0,\, T); H^{-1/2}(\partial \Omega))$$ such that
\begin{equation*}
\| A f^z_\alpha + A H g^z_\alpha \|_{ L^2( (0,\,T);\, H^{1/2}(\partial\Omega)) }< \varepsilon.
\end{equation*}
Hence, we have
\begin{eqnarray*}
&&\| F g^z_\alpha - \Gamma^0_{(y,\, s)}\|_{ L^2( (0,\, T);\, H^{1/2}(\partial\Omega)) }\\
&=& \| -AH g^z_\alpha - \Gamma^0_{(y,\, s)}\|_{ L^2( (0,\, T);\, H^{1/2}(\partial\Omega)) }\\
&\leq & \| -AH g^z_\alpha - Af^z_\alpha \|_{ L^2( (0,\, T);\, H^{1/2}(\partial\Omega)) } + \| Af^z_\alpha - \Gamma^0_{(y,\, s)}\|_{ L^2( (0,\, T);\, H^{1/2}(\partial\Omega)) }\\
&\leq & \varepsilon+\delta.
\end{eqnarray*}
To finish the proof, we recall that
$$ Hg^z_\alpha \approx f^z_\alpha \textrm{ in }L^2( (0,\, T);\, H^{-1/2}(\partial D))$$ and $$ \| f^z_\alpha \|_{L^2( (0,\, T);\, H^{-1/2}(\partial D))}\to \infty\quad (\alpha\to 0).$$
These imply
\begin{equation*}
\| Hg^z_\alpha \|_{L^2( (0,\, T);\, H^{-1/2}(\partial D))}\to \infty\quad (\alpha\to 0)
\end{equation*}
and hence
\begin{equation*}
\| g^z_\alpha \|_{L^2( (0,\, T);\, H^{-1/2}(\partial \Omega))}\to \infty\quad (\alpha\to 0).
\end{equation*}
Then, the rest of the proof can be done by noticing $\alpha=\alpha(\delta)\to 0\,(\delta\to 0)$. \hfill $\Box$

\bigskip
By Proposition \ref{prop:3} and Proposition \ref{prop:6}, we have the linear sampling method for identifying cavities in a heat conductor by measurements at its boundary.

\bigskip
{\bf Acknowledgement:} The second author is supported by Grant-in-Aid for Scientific Research (B) (22340023) of the Japan Society for the Promotion of Science. The third author is supported by NSF of China (No.11071039, 11161130002).

\bigskip

\appendix
\section{Property of potential (\ref{eq:den8})}\label{Appendix:uniqueness}
\setcounter{equation}{0}
This Appendix is devoted to showing that
$$w(x,\,t)=\int_t^T \int_{\partial D}M(y,\,s;\,x,\,t)\eta(y,\,s)ds(y)ds=0 \textrm{ in } (\mathbb{R}^n\setminus\overline D)\times (0,\,T)$$
for $w=0$ on $(\partial \Omega)_T$. It can be easily verified that
$$\partial_t w + \Delta w=0 \textrm{ in } (\mathbb{R}^n\setminus\overline{D})\times (0,\,T)$$
and $$w=0\textrm{ at }t=T.$$

Denote by $N(K^*)$ the nullspace of the operator $K^*$. Let $\eta=\eta(x,\,t)\in N(K^*)$ be such that $\eta\in C^0((\partial D)_T)$ and $\eta|_{t=0}=0$. Note that such $\eta$'s are dense in $N(K^*)$. We continuously extend $\eta$ to $t<0$ and use the same notation for the extended one.

We now establish the uniqueness of solutions to the following problem:
\begin{equation}\label{eq:B1}
\left\{\begin{array}{l}
\partial_t w + \Delta w=0 \textrm{ in } (\mathbb{R}^n\setminus\overline{\Omega})\times (-\infty,\,T),\\
w=0 \textrm{ on } \partial \Omega\times (-\infty,\,T),\\
w=0\textrm{ at }t=T.
\end{array}\right.
\end{equation}
For $s>t$ and $x\not= y$, using Theorem 6.15 in \cite{Kress} and the inequality
$$r^\beta e^{-r}\leq \beta^\beta e^{-\beta},\quad 0<r,\,\beta<+\infty,$$
we deduce that
\begin{eqnarray}\label{eq:B3}
| M(y,\,s;\,x,\,t) | &=&\left|\frac{1}{2\sqrt{4\pi(s-t)}^n} \frac{(\nu(y),\,x-y)}{s-t}\exp\left( -\frac{| x-y |^2}{4(s-t)}  \right)\right|\nonumber\\
&\leq & \frac{C_1}{\sqrt{(s-t)}^n} \frac{| x-y|^2}{s-t}\exp\left( -\frac{| x-y |^2}{4(s-t)}  \right)\nonumber\\
&\leq & \frac{C_2}{(s-t)^{\alpha_1} | x-y |^{n-2\alpha_1}}
\end{eqnarray}
for $0<\alpha_1<1/2$, where $C_1,\,C_2$ are constants. Similarly, we also have
\begin{eqnarray}\label{eq:B4}
&&|\partial_{\nu(x)}M(y,\,s;\,x,\,t)|\nonumber\\
&=&\left|\frac{1}{4\sqrt{4\pi(s-t)}^n} \frac{(\nu(y),\,x-y)}{s-t} \frac{(\nu(x),\,y-x)}{s-t} \exp\left( -\frac{| x-y |^2}{4(s-t)}  \right)\right.\nonumber\\
& &\left. + \frac{1}{2\sqrt{4\pi(s-t)}^n} \frac{(\nu(y),\,\nu(x))}{s-t}\exp\left( -\frac{| x-y |^2}{4(s-t)}  \right)\right|\nonumber\\
&\leq& \frac{C_3}{\sqrt{(s-t)}^n} \left(\frac{| x-y|^2}{s-t}\right)^2\exp\left( -\frac{| x-y |^2}{4(s-t)}  \right)\nonumber\\
& & + \frac{C_4}{\sqrt{(s-t)}^n} \frac{1}{s-t}\exp\left( -\frac{| x-y |^2}{4(s-t)}  \right)\nonumber\\
&\leq & \frac{C_5}{(s-t)^{\alpha_2} | x-y |^{n-2\alpha_2}}+\frac{C_6}{(s-t)^{\alpha_3} | x-y |^{2+n-2\alpha_3}}
\end{eqnarray}
for $0<\alpha_2,\,\alpha_3<1/2$, $s>t$ and $x\not=y$, where $C_j\,(j=3,\,\cdots,\,6)$ are constants.

Let $B_R$ be the disk with sufficiently large radius $R$ and center at the origin. Define $\Omega_R:=(\mathbb{R}^n\setminus\overline \Omega)\cap B_R$. Then, by the estimate (\ref{eq:B3}), we have
\begin{equation*}
\| w(\cdot,\,-\infty) \|_{L^2(\Omega_R)}=0.
\end{equation*}
Making use of this fact, combined with $\| w(\cdot,\,T) \|_{L^2(\Omega_R)}=0$ and  $w=0$ on $\partial \Omega\times (-\infty,\,T)$, we obtain that
\begin{eqnarray}\label{eq:B5}
0&=&\int_{\Omega_R\times (-\infty,\,T)}w (\partial_t+\Delta) w ds(y)ds\nonumber\\
&=&-\int_{\Omega_R\times (-\infty,\,T)} | \nabla_x w|^2 dyds +\int_{\partial B_R\times (-\infty,\,T)}w\partial_\nu wds(y)ds.
\end{eqnarray}
By the estimates (\ref{eq:B3}) and (\ref{eq:B4}), we can easily deduce that
$$\int_{\partial B_R\times (-\infty,\,T)}w\partial_\nu wds(y)ds\to 0 \quad  (R\to +\infty).$$
It follows from (\ref{eq:B5}) that
$$ \lim_{R\to\infty} \int_{\Omega_R \times (-\infty,\,T)} | \nabla_x w |^2 =0,$$
and therefore
$$w=0\textrm{ in } (\mathbb{R}^n\setminus\overline \Omega)\times (-\infty,\,T)$$
by observing $w=0$ on $(\partial \Omega)\times (-\infty,\,T)$.

Due to the unique continuation principle, we have
$$w=0\textrm{ in } (\mathbb{R}^n\setminus\overline D)\times (-\infty,\,T).$$
The proof is complete.

\section{The jump relation of $K^*$}\label{Appendix:jump}
\setcounter{equation}{0}

In this appendix, we show the jump relation of the adjoint of $K$ which is used in proving Lemma \ref{lem:2}. At first, we justify it for the continuous density by using the argument in the proof of Theorem 9.5 in \cite{Kress}.

We interchange the variables $(x,\,t)$ and $(y,\,s)$ in (\ref{eq:den7}), that is,
\begin{equation}\label{eq:A11}
K^*\eta(x,\,t)=\int_t^T \int_{\partial D}M(y,\,s;\,x,\,t)\eta(y,\,s)ds(y)ds
\end{equation}
with
$$ M(y,\,s;\,x,\,t)=\frac{1}{[4\pi (s-t)]^{n/2}}\partial_{\nu(y)}\exp\left( -\frac{| y-x |}{4(s-t)} \right).$$
Define
\begin{eqnarray}\label{eq:A12}
&&w(x,\,t):=(K^*\eta)(x,\,t)\nonumber\\
&=&\int_t^T \int_{\partial D} M(y,\,s;\,x,\,t)\eta(y,\,s)ds(y)ds,\, (x,\,t)\in (\mathbb{R}^n\setminus \partial{D})\times (0,\,T)
\end{eqnarray}
and set $\sigma=\frac{| x-y |}{2\sqrt{s-t}}$. Then, we have
\begin{eqnarray*}
&&w(x,\,t)\\
&=&\int_t^T \frac{1}{[4\pi (s-t)]^{n/2}} \int_{\partial D} \frac{(\nu(y),\,x-y)}{2(s-t)} \exp\left( -\frac{| y-x |}{4(s-t)} \right) \eta(y,\,s)ds(y)ds\\
&=& \frac{1}{\sqrt{\pi}^n} \int_{\partial D} \frac{(\nu(y),\,x-y)}{| x-y |^n} \int_{\frac{| x-y |}{2\sqrt{T-t}}}^{+\infty} \sigma^{n-1} \exp(-\sigma^2)\eta(y,\,t+\frac{| s-y |^2}{4\sigma^2})d\sigma ds(y).
\end{eqnarray*}
Therefore, we can view the heat potential (\ref{eq:A11}) as a harmonic double-layer potential with density
\begin{eqnarray*}
\psi(x,\,y,\,t)&:=&\frac{1}{\sqrt{\pi}^n}\int_{\frac{| x-y |}{2\sqrt{T-t}}}^{+\infty} \sigma^{n-1} \exp(-\sigma^2)\eta(y,\,t+\frac{| s-y |^2}{4\sigma^2})d\sigma\\
&=& | x-y |^n \int_t^T \frac{1}{\sqrt{4\pi(s-t)}^n} \frac{1}{2(s-t)}\exp\left( -\frac{| x-y |^2}{4(s-t)} \right)\eta(y,\,s)ds.
\end{eqnarray*}

We now show that $\psi$ is continuous on $\mathbb{R}^n\times\partial D\times [0,\,T)$ with
\begin{equation}\label{eq:A13}
\lim_{x\to y} \psi(x,\,y,\,t)=\gamma_n \eta(y,\,t)
\end{equation}
for all $y\in\partial D$ and $t\in [0,\,T)$, where
$$\gamma_n=\frac{1}{\sqrt{\pi}^n}\int_0^\infty s^{n-1}\exp(-s^2)ds.$$
To this end, we split $\psi(x,\,y,\,t)$ into three parts as follows:
\begin{eqnarray*}
&&\psi(x,\,y,\,t)\\
&=& \frac{1}{\sqrt{\pi}^n}\int_{\frac{| x-y |}{2\sqrt{T-t}}}^{\sqrt{| x-y |}} \sigma^{n-1} \exp(-\sigma^2)\eta(y,\,t+\frac{| x-y |^2}{4\sigma^2})d\sigma\\
& & + \frac{1}{\sqrt{\pi}^n}\int_{\sqrt{| x-y |}}^{+\infty} \sigma^{n-1} \exp(-\sigma^2)\left\{\eta(y,\,t+\frac{| x-y |^2}{4\sigma^2})-\eta(y,\,t)\right\}d\sigma\\
& & + \eta(y,\,t)\frac{1}{\sqrt{\pi}^n}\int_{\sqrt{| x-y |}}^{+\infty} \sigma^{n-1} \exp(-\sigma^2)d\sigma\\
&:=& I_1 + I_2 + I_3.
\end{eqnarray*}
Obviously, we have
$$\lim_{x\to y}I_1(x,\,y,\,t)=0$$ uniformly on $\partial D$ and on compact subintervals of $[0,\,T)$ and $$\lim_{x\to y}I_3(x,\,y,\,t)=\gamma_n \eta(y,\,t)$$ uniformly on $\partial D \times [0,\,T]$. Since $\eta$ is uniformly continuous, for any $\varepsilon>0$ there exists $\delta>0$ such that $|\eta(y,\,t_1)-\eta(y,\,t_2)|<\varepsilon$ for all $y\in\partial D$ and all $t_1,\,t_2\in[0,\,T]$ with $| t_1-t_2|<\delta$. Then, for all $| x-y |<4\delta$ and all $\sigma\geq \sqrt{| x-y |}$ we have
$$\left| \eta(y,\,t+\frac{| x-y |^2}{4\sigma^2})-\eta(y,\,t)  \right|<\varepsilon,$$
and therefore
$$| I_2(x,\,y,\,t)| < \frac{\varepsilon}{\sqrt{\pi}^n}\int_{\sqrt{| x-y |}} \sigma^{n-1} \exp(-\sigma^2)d\sigma\leq \gamma_n\varepsilon.$$
This means that
$$\lim_{x\to y}I_2(x,\,y,\,t)=0$$
uniformly on $\partial D\times [0,\,T]$.

Using Theorem 6.13 in \cite{Kress}, together with (\ref{eq:A13}), we can deduce that
\begin{equation}\label{eq:A14}
\lim_{h\to 0}w(x\mp h\nu(x),\,t)=\int_t^T \int_{\partial D}M(y,\,s;\,x,\,t)\eta(y,\,s)ds(y)ds \pm \frac{1}{2}\eta(x,\,t)
\end{equation}
with uniform convergence on $\partial D$ and on any compact subinterval of $[0,\,T)$. This completes the proof for the case of continuous density.

Based on the Lax theorem (\cite{Kress}), the validity in $L^2$ can be shown by using the argument of Kersten in \cite{Kersten}. The proof is complete.

\end{document}